\documentclass[conference, a4paper]{IEEEtran}
\IEEEoverridecommandlockouts
\usepackage{cite}
\usepackage{amsmath,amssymb,amsfonts}
\usepackage{algorithmic}
\usepackage{graphicx}
\usepackage{textcomp}
\usepackage{xcolor}
\usepackage{soul}
\usepackage{pdfpages}
\usepackage{hyperref}
\usepackage{tikz}
\usepackage{verbatim}
\usepackage{comment}
\usepackage{xspace}
\usepackage{subfig}
\newcommand{\wmnet}{{WM Network}\xspace}
\newcommand{\siminet}{{Similarity Network}\xspace}
\newcommand*{\rom}[1]{\expandafter\@slowromancap\romannumeral #1@}
\def\BibTeX{{\rm B\kern-.05em{\sc i\kern-.025em b}\kern-.08em
    T\kern-.1667em\lower.7ex\hbox{E}\kern-.125emX}}
\begin{document}

\makeatletter
\def\ps@IEEEtitlepagestyle{
\def\@oddfoot{\mycopyrightnotice}
\def\@evenfoot{}
}
\def\mycopyrightnotice{
{\footnotesize This paper has been accepted for publication by 2021 IEEE Visual Communications and Image Processing. The copyright is with the IEEE. \hfill} 
\gdef\mycopyrightnotice{}
}

\title{
A Deep Learning--based Audio-in-Image Watermarking Scheme\\
\thanks{
}
}

\author{



\IEEEauthorblockN{Arjon Das}
\IEEEauthorblockA{\textit{
Department of Computer Science} 
\\
\textit{University of Nebraska Omaha}
\\
arjondas@unomaha.edu
}

\and
\IEEEauthorblockN{Xin Zhong}
\IEEEauthorblockA{
\textit{Department of Computer Science} 
\\
\textit{University of Nebraska Omaha}
\\
xzhong@unomaha.edu}
}

\maketitle

\begin{abstract}
This paper presents a deep learning--based audio-in-image watermarking scheme. Audio-in-image watermarking is the process of covertly embedding and extracting audio watermarks on a cover-image. Using audio watermarks can open up possibilities for different downstream applications. For the purpose of implementing an audio-in-image watermarking that adapts to the demands of increasingly diverse situations, a neural network architecture is designed to automatically learn the watermarking process in an unsupervised manner. In addition, a similarity network is developed to recognize the audio watermarks under distortions, therefore providing robustness to the proposed method. Experimental results have shown high fidelity and robustness of the proposed blind audio-in-image watermarking scheme.
\end{abstract}

\begin{IEEEkeywords}
Audio-in-image watermarking, deep learning, neural networks, robustness
\end{IEEEkeywords}

\section{Introduction}

\begin{figure}[h!]
    \centering
    \vspace{-0.2cm}
    \includegraphics[width=0.9\linewidth]{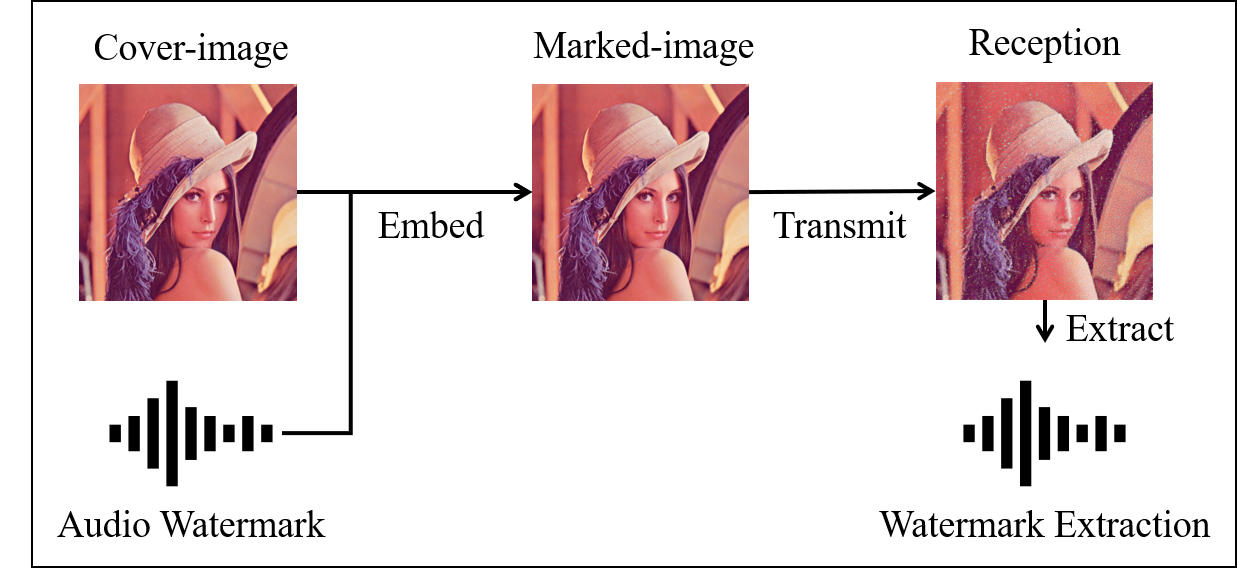}
    \vspace{-0.1cm}
    \caption{Audio-in-Image Watermarking.}
    \vspace{-0.2cm}
    \label{fig:general}
\end{figure}

Audio-in-image watermarking refers to the process of embedding and extracting audio watermark information covertly on a cover-image (see Fig.~\ref{fig:general}). An audio watermark is hidden into a cover-image to create a to-be-transmitted marked-image. The covert process suggests that the marked-image visually does not give out any watermark information, but processing the marked-image through an authorized extraction process can reveal the correct watermark. 

Based on different target application scenarios, there are many typical factors and properties a watermarking scheme should provide~\cite{9133124, 511977, 6706395}. 
For example, the fidelity that secures a high similarity between the marked-image and the cover-image, and the blindness that the watermark extractions require no information about the original images.
Compared to the highlighted properties in related areas, such as the undetectability in steganography~\cite{8305045} and the amount capacity in data hiding~\cite{10.1145/3313950.3313952}, an image watermarking scheme often considers the robustness that ensures correct watermark extraction and recognition when the marked-image is distorted or disturbed.

\noindent{\textbf{Audio watermarks:}}
From the aspect of applications, an audio watermark facilitates different downstream scenarios by providing multiple information, for example, two-factor authentication~\cite{10.1007/978-3-319-52464-1_8, 9369515} by providing the speaker and content information.
For research investigation, there is quite a number of work that separately focused on cover-image and cover-audio watermarking~\cite{1560462, 8263154, 10.5555/3184934.3287852}.
But audio-in-image watermarking remains a challenge~\cite{bhat2011audio} because audio signal spans between a limited range and inserting it into an image can cause severe tampering of its amplitudes.
Therefore, it is of great interest to explore a watermarking scheme that combines audio and image.

\noindent{\textbf{Deep learning--based image watermarking:}} Manually designing watermarking schemes often requires domain knowledge, for example, understanding a signal's frequency band to determine where to insert a specific watermark. As a result, almost every particular scenario requires a special design.
Different schemes have emerged that incorporate deep neural networks to implement image watermarking, which adapts to the demands of increasingly diverse applications.
For instance, Kandi \textit{et al.}~\cite{KANDI2017247} incorporated two convolutional autoencoders to reconstruct one cover-image so that the zero or one bits in a binary watermark stream can be indicated. Li \textit{et al.}~\cite{LI2019432} has proposed to embed a watermark into a gray-scale cover-image with manual frequency-domain methods and extract it with a convolutional neural network.
Although many methods have successfully applied deep neural networks to assist some aspects in watermarking, it has been pointed out~\cite{9133124} that designing deep learning architecture to fully learn the entire watermark embedding and extracting processes is more challenging.
Moreover, because deep neural networks can be susceptible to noises~\cite{7467366}, achieving robustness in deep learning--based watermarking is a major challenge because distorted marked-images can mislead the neural networks. Some proposals adopt adversarial training~\cite{9301856} to address the issue.
Mun \textit{et al.}~\cite{MUN2019191} simulated attacks on marked-images during training, and applied reinforcement learning to learn a robust extraction.
Zhong \textit{et al.}~\cite{9133124} trained an end-to-end neural network to learn the entire embedding and extraction and proposed an invariance layer for the robustness of common image processing attacks that modify pixel values in a marked-image.
Chen \textit{et al.}~\cite{electronics10080932} has proposed a neural network, which included attack simulations during the training, to validate if the original and the extracted watermarks are identical.

In this paper, we propose a novel deep learning--based audio-in-image watermarking scheme that achieves blindness, fidelity, and robustness simultaneously. Our major contribution is threefold. First, a deep learning framework is proposed to fully learn the watermark embedding and extracting processes for audio-in-image watermarking. Second, to reduce the requirement of watermarking domain knowledge, the proposed watermarking model can be trained in an unsupervised manner that require no manual labeling. Finally, we developed a \siminet to tolerate and recognize audio watermarks under distortions, therefore providing robustness to our scheme.



\section{
The proposed scheme
}

\begin{figure*}[ht]
    \centering
    \vspace{-0.2cm}
    \includegraphics[width=1.0\linewidth]{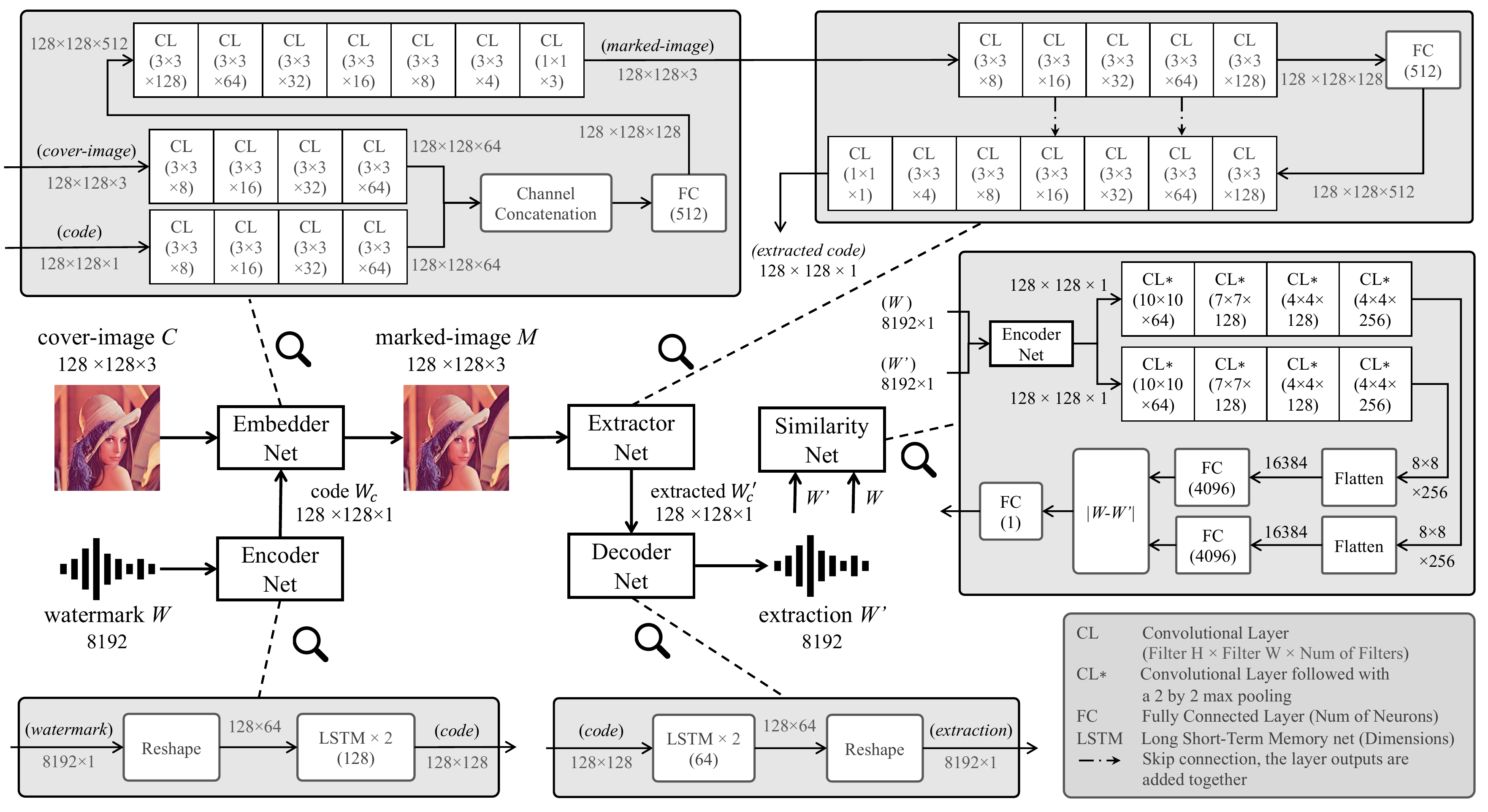}
    \vspace{-0.6cm}
    \caption{Proposed scheme: The \wmnet(Encoder, Embedder, Extractor, and Decoder Nets), and the \siminet. 
    }
    \vspace{-0.4cm}
    \label{fig:architecture}
\end{figure*}

This section describes details on (\romannumeral 1) the architecture that fully learns the process of the audio-in-image watermarking (namely the \wmnet), (\romannumeral 2) the \siminet that learns to recognize the audio watermarks under distortions, and (\romannumeral 3) the loss and training processes. The proposed scheme is summarized in Fig.~\ref{fig:architecture}.

\vspace{-0.75em}
\subsection{{\wmnet}}
Although the \wmnet is trained as a single and deep network, we conceptually modularized it to the Encoder, Decoder, Embedder, and Extractor Nets to facilitate our description and illustration.


\subsubsection{The Encoder and Decoder Nets}
The Encoder Net learns to map the watermark $W$ to its code $W_c$: $W \rightarrow W_c$ from the training samples $w_i \in W, i = 1, 2, 3, ...$.
Inversely, the Decoder Net learns to map the extracted code $W'_c$ to the watermark extraction $W'$: $W'_c \rightarrow W'$.
The Encoder Net is a two layer LSTM. The audio input is firstly reshaped from $8,192\times1$ to $128\times64$.
The first LSTM takes a $64$-dimensional input vector per time-step, and both LSTM layers are configured to have $128$-dimensional hidden state vectors. Hence the output from each LSTM cell is a $128$-dimensional vector.
After $128$ time-steps, the output size will be $128\times128$.
Symmetrically, the Decoder Net converts back a $128\times128\times1$ feature map into a $8,192\times1$ audio data.

\subsubsection{The Embedder and the Extractor Nets}
The Embedder Net learns a function that includes the $128\times128\times1$ $W_c$ into the $128\times128\times3$ cover-image $C$ while retaining visual fidelity.
First, $W_c$ and $C$ are separately processed by two sequences of convolutional layers to produce $128\times128\times64$ feature maps.
Each sequence contains four convolutional layers of $8$, $16$, $32$, and $64$ filters respectively.
Although identically structured, these sequences do not share any weights because they are processing image and audio separately.
Then, the feature maps of both $W_c$ and $C$ are concatenated along the channel dimension to obtain a $128\times128\times128$ feature map.
To promote the propagation of information throughout the feature map channels, a channel-wise fully connected layer~\cite{pathakCVPR16context} with $512$ units is applied.
The output ($128\times128\times512$) is then fed into another sequence of convolutional layers with $128$, $64$, $32$, $16$, $8$, $4$, and $3$ filters to obtain a marked-image $M$.

The Extractor Net concentrates on distilling out the watermark information embedded inside the marked-image. 
To achieve an $M \rightarrow W_c'$ mapping, the Extractor Net takes input $M$ and passes it through sequences of convolutional layers with different filter numbers. 
A $512$-unit channel-wise fully-connected layer is also applied between two sequences of convolutional layers. 
The final convolutional layer of the extractor net outputs a $128\times128\times1$ $W_c'$. To ensure successful gradient propagation throughout the extractor net, we incorporated two skip connections~\cite{7780459}.

\vspace{-0.5em}
\subsection{\siminet}

Marked-images may distort during transmission. Consequently, the watermark extractions might get distorted. 
Although experimentations reveal that the proposed \wmnet is robust to some noises, it is necessary to identify the authenticity of the distorted watermark extractions.
Simple error (e.g., RMSE) analysis fails to indicate such a validation process. 
Hence, we introduce the \siminet to validate if a watermark $W$ and its extraction $W'$ are similar by structural analysis of audio signal pairs. 
The ideal training result is to be consistent with human judgment. Explicitly, if an audio clip is distorted but recognizable by humans, it should also be recognizable by the \siminet. 
As a result, from extraction to validation, the \siminet provides robustness from a semantic recognition perspective.


The \siminet takes a pair of $8,192\times1$ ($W$, $W'$) as input and outputs a [0,1] scalar where 0 means completely different and 1 means identical, and values in between indicating a similarity level.
We formulate this similarity computation as metric learning and develop our \siminet based on the Siamese architecture~\cite{koch2015siamese}.
Our \siminet firstly applies our trained Encoder Net so that the pre-trained LSTM layers can extract features from the audio inputs rapidly.
After obtaining the $128\times128\times1$ feature maps of $W$ and $W'$, two identical weight-shared sequences of convolutional layers with max-pooling is applied to compress the feature maps to $8\times8\times256$. These feature maps are then flattened to $16,384$ and passed to a $4,096$-unit fully-connected layer.
The absolute difference is computed between the two $4,096$ feature vectors of $W$ and $W'$, and a single-unit fully-connected layer with Sigmoid activation is used to output the similarity scalar.

\vspace{-0.5em}
\subsection{Training and Loss Functions}
\label{sec:training}

\subsubsection{Encoder-Decoder Net Pre-training}
While the marked-image is output at the middle of the \wmnet, the watermark extraction has to propagate through the entire \wmnet.
As a result, the extraction may fail to pass through the rest of the network during the training.
To address this issue, we pre-train the Encoder and Decoder Nets, reconstructing $W$ to itself.
Hence, the first step of pre-training the proposed scheme is to train an Encoder-Decoder net, thus inputting $W$ to the Encoder, and the output of the Decoder should be $W$ itself.
Given training sample watermarks $w_i \in W, i = 1,2,3, ...$, the loss function for training this Encoder-Decoder net is a simple mean squared error between $w_i$ and its reconstruction $\bar{w_i}$:

\vspace{-1em}
\begin{equation}
    \mathcal{L}(w_i, \bar{w_i}) = {(w_i - \bar{w_i})}^2.
    \label{eq:pre_training_loss_function}
\end{equation}

\vspace{-0.5em}
\subsubsection{Training the \wmnet}
The second step is to train the entire \wmnet as a single network where the (pre-trained) Encoder and Decoder Nets are fine-tuned, and the Embedder and Extractor Nets are trained from scratch. Given the samples $w_i \in W, i = 1,2,3, ...$ and $c_i \in C, i = 1,2,3, ...$, an extraction loss that minimizes the difference between the extraction $w_i'$ and $w_i$ is applied to ensure a successful extraction of the watermark. In addition, a fidelity loss that minimizes the difference between a marked-image $m_i$ and $c_i$ is applied to enable a high fidelity. Mean squared errors are applied to compute the differences, and the loss is given as:

\vspace{-0.4cm}
\begin{equation}
    \mathcal{L}(w_i, w_i', c_i, m_i) = \lambda_1 {(w_i - w_i')}^2 + \lambda_2 {(c_i - m_i)}^2,
    \label{eq:wmnet_loss_function}
\end{equation}
where $\lambda_1$ and $\lambda_2$ are the weighing factors.

\subsubsection{Training the \siminet}
Given the input pair $w_i^1$ and $w_i^2$, \siminet is trained to recognize if they are the same. Let $y(w_i^1, w_i^2)$ be the label, then $y(w_i^1, w_i^2)=1$ if $w_i^1$ and $w_i^2$ are the same, and $y(w_1, w_2)=0$ otherwise. 
The network prediction $p$ is a decimal between 0 to 1, and we applied the binary cross-entropy to compute the loss:

\vspace{-0.1cm}
\begin{equation}
\begin{aligned}
    \mathcal{L}(w_i^1, w_i^2) = y(w_i^1, w_i^2)log(p) + \\ (1- y(w_i^1, w_i^2))log(1-p).
    \label{eq:bce_loss_function}
\end{aligned}
\end{equation}

\section{Experiments and Analysis}

\vspace{-0.5em}
To the best of our knowledge, our proposal is the first deep learning--based audio-in-image watermarking scheme, so it is difficult to find peer methods for a fair analogous study. We are only able to find a related handcrafted audio-in-image method~\cite{bhat2011audio} that has reported a Root Mean Square Error (RMSE) of $0.022325$ averaged over four audio watermarks, whereas our proposed scheme has an RMSE of $0.009452$ averaged over $5,800$ different audio watermarks. To further illustrate the effectiveness of the proposed scheme, we have conducted typical experiments and analyses in deep learning and image watermarking.

\vspace{-0.5em}
\subsection{Data Preparation}

To train the proposed \wmnet, we used rescaled $128\times128$ Microsoft COCO Dataset~\cite{10.1007/978-3-319-10602-1_48} as the cover-images and resampled ($8,192$ sampling rate) Speech Commands Dataset~\cite{speechcommandsv2} voice commands as watermark audio. Combining these two datasets, we have sampled $42,600$, $9,700$, and $5,800$ audio-image pairs for training, validation, and testing. None of the training audio commands match with either validation or testing audio commands.

Speech Command Dataset samples are paired with extracted audio samples derived from the pre-trained \wmnet with varying marked-image distortions to prepare the \siminet training dataset. This allowed us to generate $340,000$, $64,000$, and $46,000$ audio-audio pairs for training, validation, and testing. The audio pairs are labeled as $0$ or $1$, where $0$ means an audio pair is different and $1$ means the same.


\vspace{-0.5em}
\subsection{Training, Validation, and Testing Results}

\begin{figure}
    \vspace{-0.35cm}
    \centering
    \subfloat[]{\includegraphics[width=0.485\linewidth]{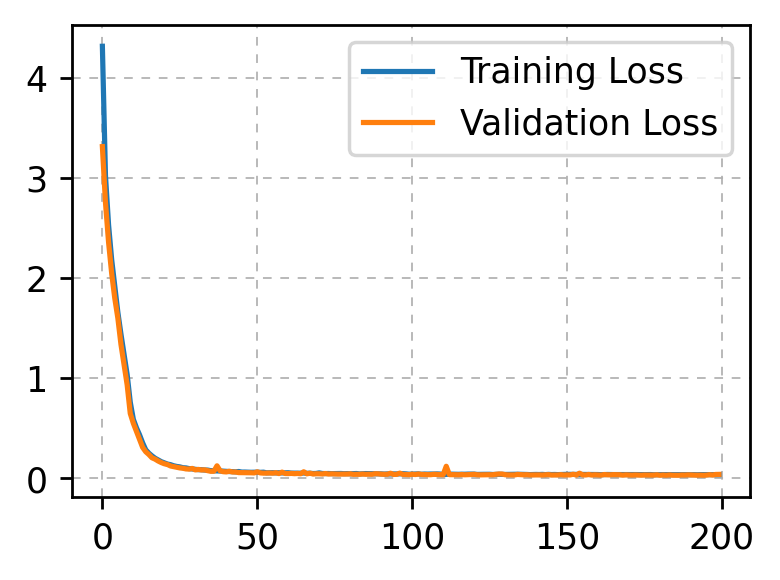} \label{fig:loss_vs_epoch}}
    \subfloat[]{\includegraphics[width=0.485\linewidth]{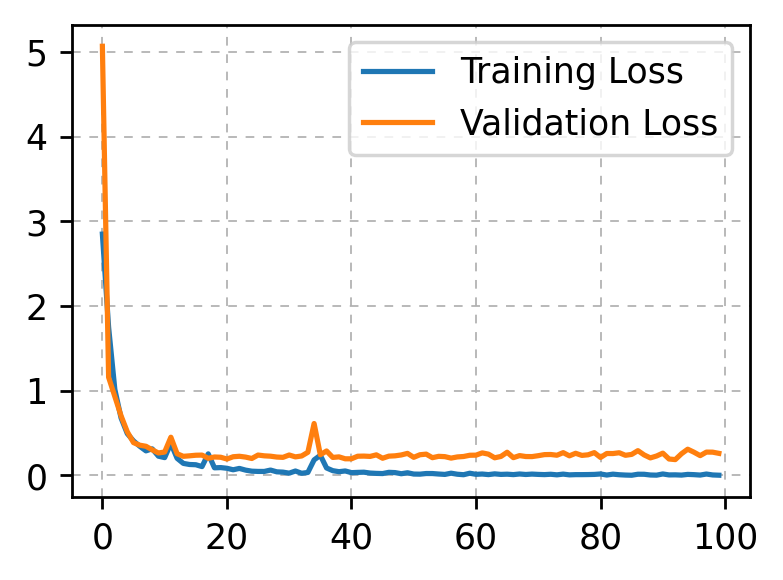}
    \label{fig:similarity_loss_vs_epoch}}
    \vspace{-0.05cm}
    \caption{(a) Learning Curves of the \wmnet for 200 epochs, and (b) Learning Curve of the \siminet for 100 epochs.}
    \vspace{-0.5cm}
\end{figure}

After preparing the datasets, we train the proposed scheme according to the strategies described in Section~\ref{sec:training}.
Fig.~\ref{fig:loss_vs_epoch} shows the loss values of training and validation of the \wmnet, where we can observe a proper fit for hundreds of epochs.
Fig.~\ref{fig:similarity_loss_vs_epoch} shows the training and validation loss of the \siminet for 100 epochs. After training, the model achieved a validation accuracy of $99.44\%$ on identifying whether a watermark, watermark extraction pair ($W$, $W'$) is the same or not.

\begin{figure}[h!]
    \centering
    \vspace{-0.3cm}
    \includegraphics[width=0.85\linewidth]{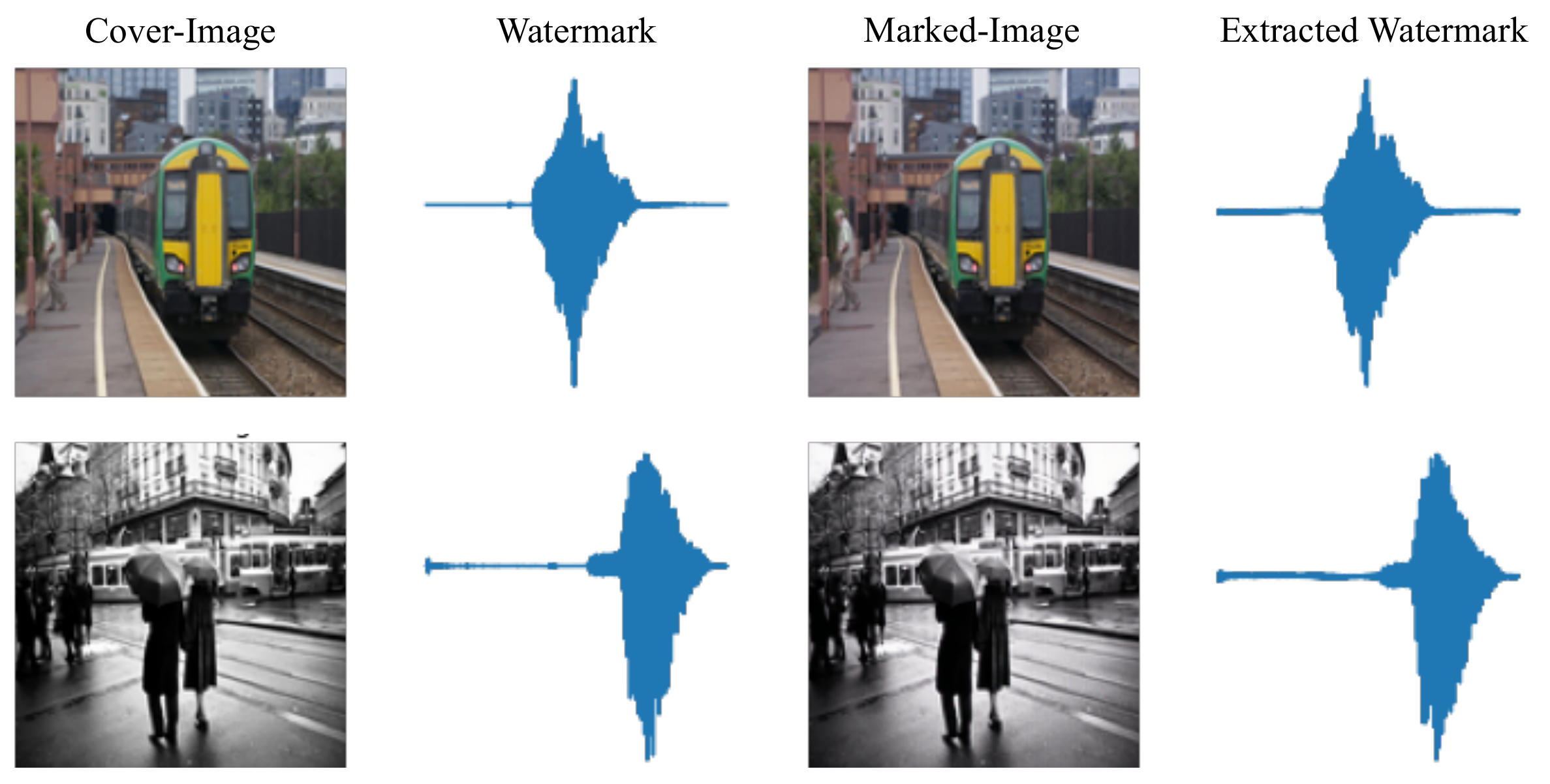}
    \vspace{-0.1cm}
    \caption{\wmnet example inputs and outputs: (from left to right) cover-images, watermarks, marked-images, and watermark extractions.}
    \vspace{-0.12cm}
    \label{fig:correctness_fig}
\end{figure}

At the testing phase of the \wmnet, we calculated the extracted watermark's RMSE and the marked-image's Structural Similarity Index Measure (SSIM)~\cite{1284395} score with the original watermark and cover-image, respectively.
At validation and testing, the watermark's RMSEs are $0.010664$ and $0.009452$, and the marked-image's SSIMs are $0.988365$ and $0.988230$, where we can observe successful extractions and high fidelity.
Fig. \ref{fig:correctness_fig} shows examples of cover-images, watermarks, marked-images, and extractions.
To evaluate the \siminet, we used binary accuracy and the cross-entropy loss value.
At validation and testing, the cross-entropy values are $0.29$ and $0.42$, and the accuracies are $99.33\%$ and $98.98\%$. Thus, the \siminet has successfully learned to recognize the watermarks even under noisy situations.

\vspace{-1.2em}
\subsection{Robustness}
The encoder introduces some decomposition and redundancy by enlarging the $8,192$ lengths $128\times64$ watermark to a $128\times128$ code. As a result, the \wmnet can resist noise to an extent without presenting any adversarial examples during the training.
Moreover, because our watermark extraction is verified by the \siminet, which can tolerate some distortions, we analyzed the overall scheme's (\wmnet + \siminet) robustness by plotting the accuracy of the \siminet versus swept-over noise parameters. 
To train the \siminet, we generate noisy audio samples derived from \wmnet with marked images where $0$ to $50\%$ cutout (randomly removing regions) or $-5^{\circ}$ to $+5^{\circ}$ rotation is applied. 
We then tested the \siminet with watermark pairs within and beyond the range of its training noise parameters. E.g., in terms of cutout noise, we tested ($W$, $W'$) pairs for up to $90\%$ noise exposure on the marked-images. 
Fig.~\ref{fig:sia_vs_cutout} shows the accuracy of \siminet versus up to $90\%$ cutout, and Fig.~\ref{fig:sia_vs_rotate} shows the accuracy versus $-6^{\circ}$ to $+6^{\circ}$ rotation on marked-images. On extracted watermark pair validation task, the model has a high and decent tolerance range against image processing and geometric rotation attacks, respectively.


\vspace{-0.5em}
\subsection{Ablation Study of the \siminet}
\begin{figure}[h!]
    \centering
    \vspace{-0.2cm}
    \includegraphics[width=0.85\linewidth]{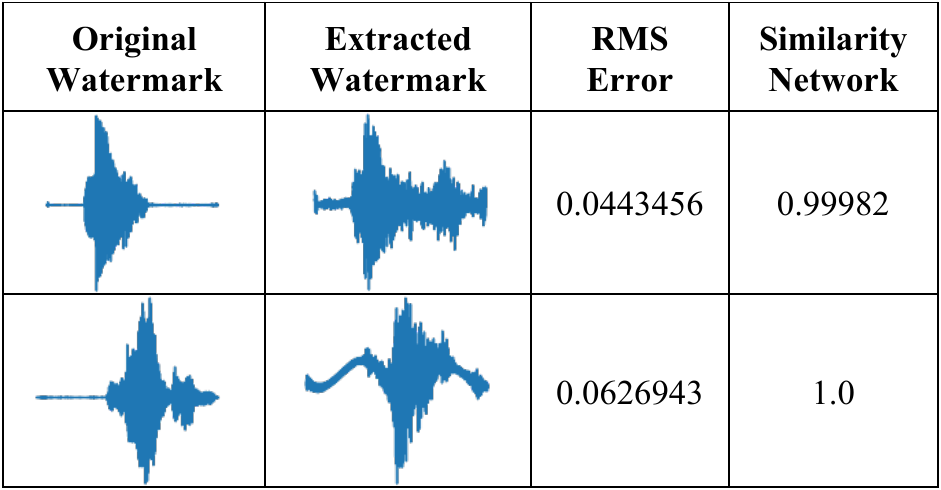}
    \vspace{-0.1cm}
    \caption{Examples of watermark and noisy extractions with RMSE and Similarity Net prediction.}
    \vspace{-0.5cm}
    \label{fig:similarity_ablation}
\end{figure}

This section demonstrates the indispensability of the \siminet by ablation study.
If there are some noises on the marked-image, the watermark extraction can be noisy while retaining some resemblance with the original.
However, simple error values cannot accurately indicate this similarity.
Fig.~\ref{fig:similarity_ablation} shows two such instances with ($W$, $W'$) pairs, and each pair is audibly the same for human hearing despite having high RMSE values.
Analytically, these values incorrectly indicate low similarity compared to the test set's low average RMSE of $0.009452$ (considering as baseline).
In contrast, the \siminet correctly outputs high similarity values.

\begin{figure}
    \vspace{-2.5em}
    \centering
    \subfloat[]{\includegraphics[width=0.485\linewidth]{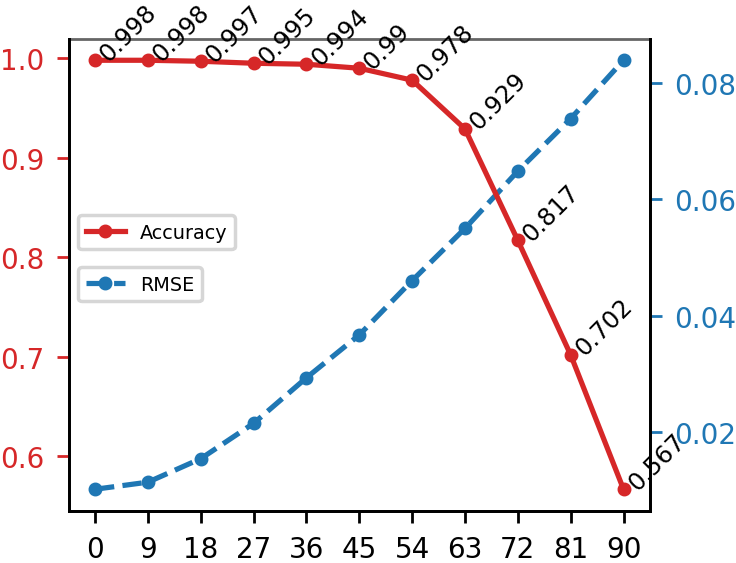} \label{fig:sia_vs_cutout}}
    \subfloat[]{\includegraphics[width=0.485\linewidth]{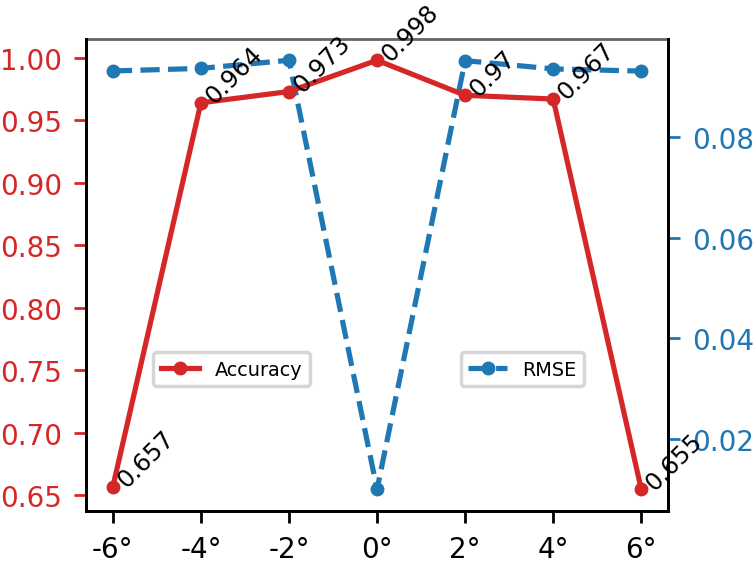} \label{fig:sia_vs_rotate}}
    \vspace{-0.1cm}
    \caption{Distortion parameters versus the \siminet accuracy and watermark RMSE, (a): \% of cutout and (b): rotation degree.}
    \vspace{-0.1cm}
\end{figure}

For a large-scale experiment on the testing set, Fig.~\ref{fig:sia_vs_cutout} and \ref{fig:sia_vs_rotate} show the RMSE and the \siminet accuracy versus the distortion parameters, where we find that under noise, the \siminet can capture the semantic similarity that is consistent with human judgments while the RMSE values are simply computing the differences.

\section{Conclusion}
This paper introduces a robust and blind audio-in-image watermarking scheme with deep learning. A \wmnet is designed to fully learn the audio-in-image watermarking process without human supervision, and a \siminet is developed to enhance robustness from the semantic recognition perspective. 
In the future, our primary goal is to improve the robustness of the \wmnet itself so that our proposed scheme can be more resistant to different attacks, especially geometric distortions.

\bibliography{references}
\bibliographystyle{IEEEtran}


\end{document}